# Experience in applying remote technology in the secondary education institutions in Russia, located in rural areas (From the experience of Podolsky municipal district schools)

Kochetkova Nina Alexandrovna, deputy principal of the MEI "Lyceum № 1".

*Abstract*
This article examines the experience of distance education technologies in Podolsky municipal district, Moscow region.

During the past three decades distance education has become a global phenomenon of educational and informational cultures, changing the face of education in many countries around the world. A whole industry of educational services has emerged and is rapidly developing, united on general name of "distance education", an impressive quantity of students, educational institutions, its size, complexity, infrastructure and scale of investment. Development of distance education is recognized as one of the key areas of basic education programs of UNESCO "Education for all", "Education throughout life," "Education without borders".

Long-term goal of distance education in the world is to enable every student living in any place, take a course of any educational institution. This presupposes a transition from the concept of the physical movement of students to the concept of mobile ideas, knowledge and education in order to spread knowledge through the sharing of educational resources.

Increased use of a distance education leads to significant improvements in access to education. For example, people with disabilities can receive distance education. It should be noted that the current level of development of remote technologies allowed to approach the efficiency of education to the traditional full time tuition, and in some cases to surpass it.

For the introduction of core education in rural schools it is often not enough available staff and academic resources. Distance core education using remote resources and teachers - one of the most effective ways of solving these problems.

A significant factor in the development of distance education system is its profitability. The average estimation shows that it is cheaper than traditional forms of education. This is achieved, on the one hand, through the effective use of educational areas, technical devices, vehicles, on the other – due to concentrated and unified presentation of educational information and multi-access to it. The study of scientific and theoretical sources (works of A.A. Andreev, E.S. Polat, V.P. Tikhomirov, A.V. Tikhonov and others) proves that distance education technologies allow us to solve a number of important educational objectives:
• creating of a single educational space;
• building students' cognitive independence and activity;
• development of critical thinking, tolerance and willingness to constructively discuss different points of view.

Experiment in order to introduce distance education into the educational process was conducted in Podolsky municipal district schools for several years. During this period, the work involved teachers and students of educational institutions of the region. As a virtual learning environment, posted on the Internet, Moodle publicly available environment has been selected.

A series of remote seminars for educators was conducted, which attracted more than 200 people, including rural schools teachers, deputy school principals, principals, teaching methods and management specialists.

Practice has shown that remote form of seminars is very relevant and in demand in the pedagogical community. The survey regarding the need for D.E. was conducted among the participants. Here are some positive D.E. aspects, which were mentioned by teachers: now there's no need to spend a lot of time and money for travel expenses to attend the required seminar, it also removes a psychological barrier when discussing any issue in the forum, you can take part in remote seminar and express your opinion in the forum, it can be done at any time and from any computer connected to the Internet (for example at home) etc. Also, except for planned organizational issues, certain topics appeared in a virtual environment, which were offered for discussion by the participants themselves, this is a very important aspect of the discussions. It turns out to be some kind of a remote "round table". In addition to discussions teachers posted their articles on the topic, presentations, and their students' works. Use of a modern internet technology and distance education allows us to easily form different virtual professional communities (e.g. teachers community) where they can communicate with each other, discuss problems, solve common issues, share information, etc., that is also a kind of advanced training of teachers through the direct exchange of experiences with each other.

Another form of work is involvement and organization of seminars with the video broadcasting over the Internet. With the help of well-known programs for video conference Skype, Videoport and others, teachers and students of educational institutions of the region took part in the regional, All-Russian and international scientific conferences, seminars and webinars. Their own online activities were also organized, including presentation of lyceum anthology "Under-Wood" with virtual participation of the Podolsk district mayor, and Chancellor of the Institute of Open Education and others.

Distance education "course of a beginner journalist" was also organized. Students aged 14-17 from various Podolsky district schools were trained in basic journalism with a practical work, creative jobs, practiced in writing essays, notes and articles.

Remote district games and competitions were also conducted: the geography, ethnography, history, etc.

Development of distance education in the Russian education system will continue with the development of internet technology and improvement of distance education methods which favor mass distribution of education, making it more affordable than traditional one. Further development of distance education systems involves ensuring maximum interactivity. It's not a secret that learning process becomes complete only when the simulation of a real communication with the teacher is reached, that's exactly what we should strive for. You must use a combination of different types of electronic communications, to compensate for lack of a personal contact through virtual communication.

The main advantage of distance education is that it allows to realize two fundamental principles of modern Education - "education for all" and "education throughout life."